\documentclass{raa}           
\usepackage{graphicx,times}
\usepackage{natbib}
\usepackage{amssymb,amsmath}
\usepackage[T1]{fontenc}
\usepackage{ae,aecompl}
\usepackage{multirow}   % plot tables
\usepackage{gensymb}    % degree symbol
\bibpunct{(}{)}{;}{a}{}{,}

\newcommand{\HI}{\hbox{\rmfamily H\,{\textsc i}}}
\newcommand{\HIsub}{\hbox{{\scriptsize H}\,{\tiny I}}}
\newcommand{\OHI}{\hbox{$\Omega_{\HIsub}$}}

\usepackage[a4paper=true,pagebackref=true]{hyperref}
\hypersetup{colorlinks = true, linkcolor = green, anchorcolor = red, citecolor = blue, filecolor = red, pagecolor = red, urlcolor = red}

\begin{document}
\title{An HI intensity mapping survey with a Phased Array Feed}
 \volnopage{ {\bf 20XX} Vol.\ {\bf X} No. {\bf XX}, 000--000}
   \setcounter{page}{1}

\author{Lincheng Li\inst{1,2,3,4}, 
   Lister Staveley-Smith\inst{3,4,5}, 
   Jonghwan Rhee\inst{3,4,5} 
   }
 
\institute{National Astronomical Observatories, Chinese Academy of Sciences, Beijing 100012, China; {\it lilincheng@bao.ac.cn}\\
\and
School of Astronomy and Space Science, University of Chinese Academy of Sciences, Beijing 100049, China\\
\and
International Centre for Radio Astronomy Research (ICRAR), The University of Western Australia, 35 Stirling Hwy, Crawley, WA, 6009, Australia\\
\and 
ARC Centre of Excellence for All-sky Astrophysics (CAASTRO)\\
\and 
ARC Centre of Excellence for All Sky Astrophysics in 3 Dimensions (ASTRO 3D)\\
\vs \no
   {\small Received 20XX Month Day; accepted 20XX Month Day}
}

\abstract{We report results from a neutral hydrogen ({\HI}) intensity mapping survey conducted with a Phased Array Feed (PAF) on the Parkes telescope. The survey was designed to cover $\sim380$~deg$^2$ over the redshift range $0.3 < z < 1$ (a volume of $\sim 1.5$~Gpc$^3$) in four fields covered by the WiggleZ Dark Energy Survey. The results presented here target a narrow redshift range of 0.73 $< z <$ 0.78 where the effect of radio frequency interference (RFI) was less problematic. The data reduction and simulation pipeline is described, with an emphasis on flagging of RFI and correction for signal loss in the data reduction process, particularly due to the foreground subtraction methodology.  A cross-correlation signal was detected between the {\HI} intensity maps and the WiggleZ redshift data, with a mean amplitude of 
$\left\langle\Delta T_\textrm{b}\delta_\textrm{opt}\right\rangle = 1.32\pm 0.42$~mK (statistical errors only).
A future Parkes cryogenic PAF is expected to detect the cross-correlation signal with higher accuracy than previously possible and allow measurement of the cosmic \HI\ density at redshifts up to unity.
}

   \authorrunning{L. Li et al. }            %author_head in even pages
   \titlerunning{Phased Array Feed HI Intensity Mapping}  % title_head in odd pages
   \maketitle

%________________________________________________ sections below
% 
\section{Introduction}           %% first-level sections will be auto-capitalized
\label{sect:intro}
The information of matter distribution of the universe is imprinted on the Large Scale Structure (LSS) traced by galaxies. So measurements of LSS provide essential constraints to cosmological models.
Previously, most LSS studies have been made based on optical redshift and lensing surveys such as 2dFGS \citep{Colless01}, SDSS \citep{York00}, WiggleZ \citep{WigglZ}, BOSS \citep{Anderson12} and DES \citep{Flaugher05}. Planned future optical surveys include LSST \citep{LSST}, Euclid \citep{Euclid} and DESI \citep{DESI}.
However, alternative techniques are also available through radio continuum and spectral-line surveys, particularly with the future Square Kilometre Array (SKA) and its pathfinders \citep[e.g.][]{Prandoni15,EMU,Staveley-Smith:2015,DINGO,WALLABY}. These surveys have the potential to detect all massive galaxies at $z<2$, and therefore to measure LSS with minimal cosmic variance.

At the current time, the ability to detect individual galaxies in neutral hydrogen (\HI) at cosmological distances is limited because of the weakness of the 21-cm emission line with which the mass of \HI\ in galaxies is measured. The most distant galaxy so far detected in the 21-cm \HI\ emission line is at $z=0.37$ detected in the COSMOS \HI\ Large Extragalactic Survey (CHILES) using Karl G. Jansky Very Large Array (JVLA) \citep{Fernandez16}. However, the techniques of `stacking' and `intensity mapping' (IM), whereby statistical measurements are made by summing the emission from a large sample of galaxies \citep{Rhee16,Rhee18}, or by measuring the correlation function in 21-cm images \citep{Chang10, Masui2013}, can in principle allow the detection of LSS at redshifts of about unity, even with current radio telescopes. Telescopes such as 
BAO from Integrated Neutral Gas Observations \citep[BINGO,][]{Battye12}, Canadian Hydrogen Intensity Mapping Experiment \citep[CHIME,][]{CHIME}, Hydrogen Intensity and Real-time Analysis eXperiment \citep[HIRAX,][]{Newburgh16} and Tian-Lai \citep{Chen12} are currently being constructed with the aim of measuring LSS with high accuracy via \HI\ intensity mapping. We refer the reader to \citet{Bull15} for a comprehensive introduction to intensity mapping.

In addition to the study of LSS, a further application for IM is to measure the cosmic neutral hydrogen mass density ({\OHI}) at intermediate redshifts. 
At $1.5<z<5$, {\OHI} has been estimated using Damped Lyman-$\alpha$ Absorption (DLA) systems which are defined by deep absorption lines in the spectra of QSOs, due to intervening clouds of hydrogen \citep[e.g.][]{Noterdaeme2012,Crighton15,Bird17}. 
At lower redshifts of $z<1.5$, Lyman-$\alpha$ absorption lines are shifted to ultraviolet wavelengths and become undetectable using ground-based telescopes. 
However, at these intermediate redshifts, by cross-correlation between the {\HI} density field and the optical density field in regions of the sky with suitable deep redshift surveys, IM can be used to infer the amplitude of fluctuations in the {\HI} density field, and thereby the cosmic {\HI} mass density {\OHI}.

Several pioneering experiments in intensity mapping have been undertaken.
\citet{Pen08} reported a convincing cross-correlation signal between \HI\ Parkes All Sky Survey \citep[HIPASS,][]{Barnes01} data and optical data from the 6dF Galaxy Survey \citep[6dFGS,][]{Jones04,Jones09}.
\citet{Chang10} reported the first detection of a signal at the cosmologically significant redshift $z\sim 0.8$ using the Green Bank Telescope (GBT), which was followed by \citet{Masui2013} and \citet{Switzer13}. \citet{Anderson18} then explored the effect of environment on \HI\ content by cross-correlating Parkes data with the 2dFGRS strip across the South Galactic Pole.
The first detection of the power spectrum via a interferometer was obtained by \citet{Vujanovic} with the Australia Telescope Compact Array. Furthermore \cite{Santos15} and others have pointed out that the upcoming Square Kilometer Array (SKA) will be powerful in conducting future IM surveys. 

However, many challenges remain in IM experiments. 
The first is the thermal noise from the telescope which, at $z\sim$ 1 is predicted to be 5 $\sim$ 6 orders of magnitude greater than the predicted IM temperature fluctuations. The second is contamination from Galactic and extragalactic foregrounds which are $\sim$ 4 orders of magnitude larger than the {\HI} signal, and significantly polarised. Finally, radio frequency interference (RFI) from terrestrial sources is prevalent at low frequencies, and can easily exceed any other source of noise by many orders of magnitude.
Thermal noise is uncorrelated with time, and can be suppressed in the normal manner by integration over time and sampling a wide range of frequencies. Foregrounds mainly consist of the diffuse synchrotron and free-free emission from Galactic and extragalactic sources, which are featureless and can therefore be removed by smoothly varying functions if accurate bandpass calibration is available \citep{Condon92}. More sophisticated foreground subtraction techniques have also been developed for dealing with IM data and similar data from EOR experiments \citep{Wang06,Liu2011,Chapman12,Switzer13,Wolz15}. RFI is normally mitigated by flagging and removal in spectral space, although sophisticated adaptive filtering techniques can also provide additional suppression \citep{Briggs17,Reynolds17}.

In this paper, we introduce a new IM survey aiming to measure the large scale {\HI} emission in the redshift range $0.3<z<1$.
This survey was conducted with Parkes telescope equipped with a novel Phased Array Feed (PAF) \citep{Chippendale16}, which offers several advantages over normal telescope feeds \citep{Reynolds17,Deng17}. 
The most important advantages of the PAF for the present work are the ability to form multiple simultaneous beams, to cover a large bandwidth, and to present a well-calibrated bandpass without problems associated with spectral ripple. We cross-correlated the measured {\HI} density field with optical density field derived from the catalogues of WiggleZ Dark Energy Survey and detected the cross-correlation signals at redshift $z\sim0.75$.

The paper is structured as follows: in Section 2 we introduce our observational strategy and data reduction pipeline in detail. In Section 3 we verify the pipeline using simulations. Section 4 details the foreground removal method that we used to alleviate residual foreground contamination. The final results and estimated cosmological parameters are reported in Section 5, and we summarise in section 6.
Throughout, the cosmological parameters are given by \citet{Komatsu09}, with $\Omega_mh^{2}=0.1358$, $\Omega_b=0.0456$, $\Omega_\Lambda=0.726$, and $h=0.705$.

% Authors can give a citation as `\citealt{Michel+etal+1992}'.
% You may also use \cite, \citep and \citet for citation, and use Table~1
% or Figure~1 and so forth. Using \ref and \label for cross-references of
% Tables/Figures is a good way in adjusting/adding/removing text, tables or
% figures.

\section{DATA}
\subsection{HI Observations}
\label{sec:observation} 

\begin{figure}
    \centering
	\includegraphics[width=0.8\columnwidth]{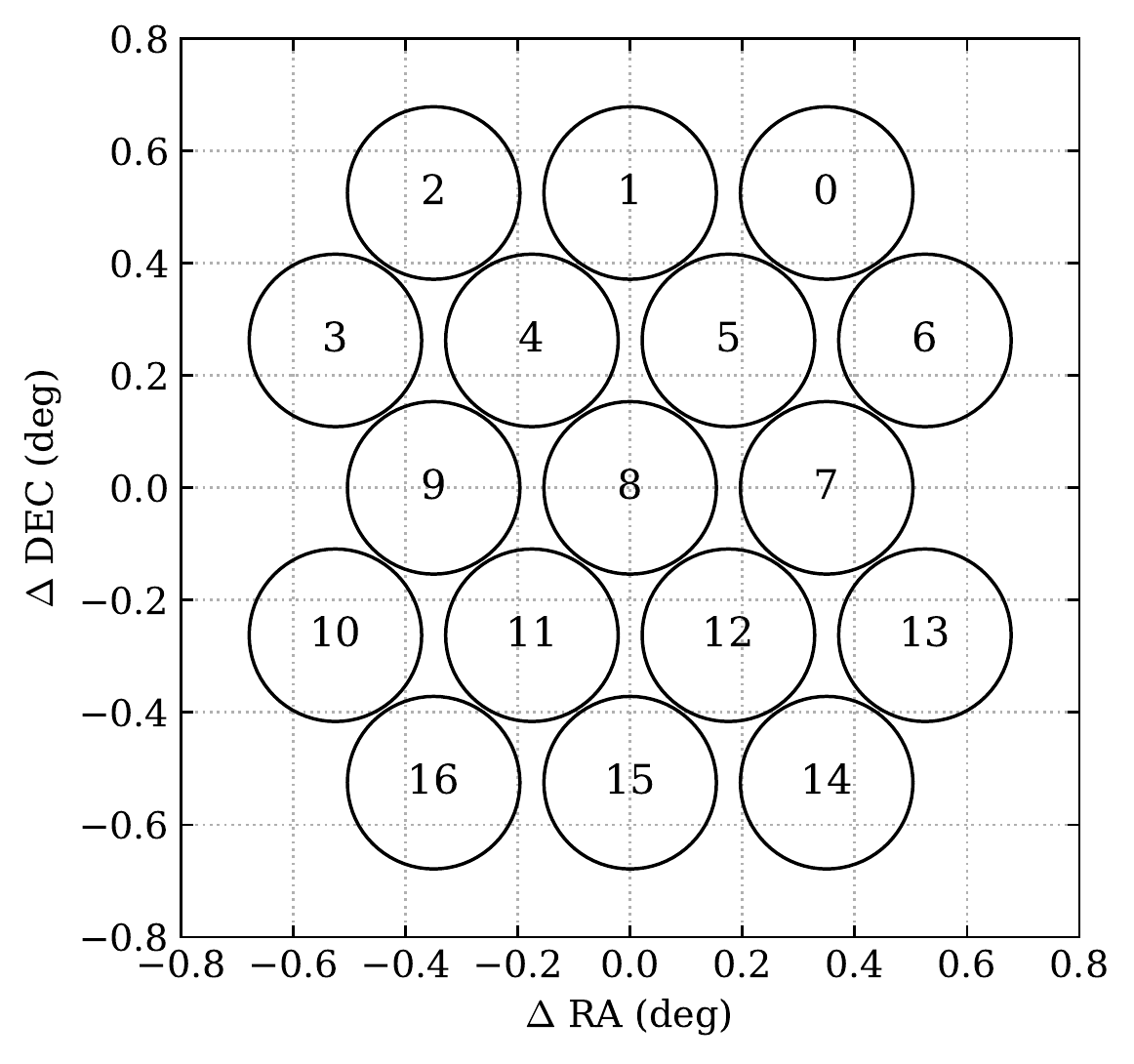}
    \caption{The 17-beam footprint of the phased array feed used in the observation.}
    \label{17beams}
\end{figure}

\begin{figure}
    \centering
	\includegraphics[width=\textwidth]{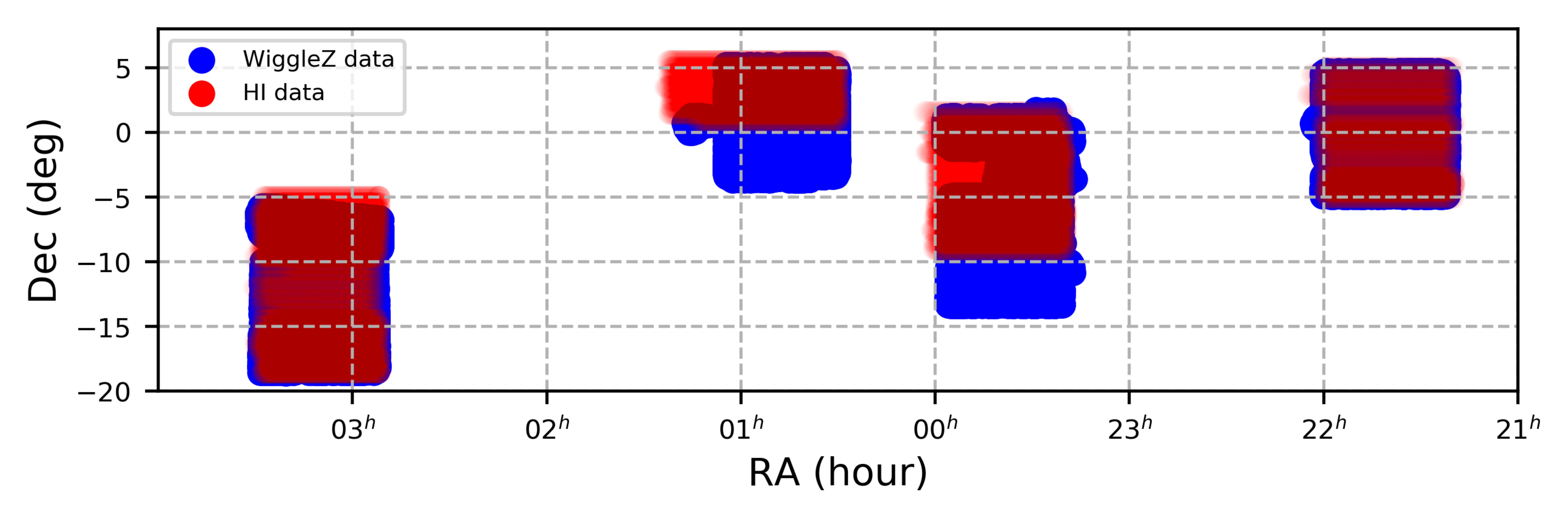}
    \caption{The Parkes PAF intensity mapping survey coverage (red), overlaid on the WiggleZ dark energy survey footprint (blue).}
    \label{footprints}
\end{figure}

The \HI\ data were obtained from observing program P913 (PI: Staveley-Smith), taken using the CSIRO Parkes 64-m telescope
%\footnote{The Parkes radio telescope is part of the Australia Telescope National Facility which is funded by the Australian Government for operation as a National Facility managed by CSIRO.} 
in 2016 September and October.  
The observations were conducted using CSIRO-built PAF for the Max-Planck-Institute for Radio Astronomy (MPIfR), which was commissioned at Parkes prior to delivery to the Efflesberg 100-m telescope (see \citet{Chippendale16} for more details). 
The data used here were taken in the so-called band 1, which covers a frequency range of 700 to 1084 MHz. The surveyed regions were chosen to coincide with the footprints of the 0h, 1h, 3h and 22h fields of the WiggleZ survey \citep{WigglZ}. 
During the observation a beamformer was used to form 17 discrete beams, each with a measured half-power width of $22.3^\prime$ at the central frequency we considered in this paper.
Due to firmware limitations only 16 beams could be channelised. The beam spacing pitch was $0.35^{\circ}$ and, for the purposes of analysis, the beam shape is assumed to be Gaussian. 
Fig.~\ref{17beams} shows the footprint of the phased array feed used in the observations.

The flux density calibrators used for the observations were PKS~B1934-638 and Hydra~A, depending on the RA of the targeted fields during the observation. To maximise the stability of the data, and to minimise the effect of sidelobe and RFI variation, the science observations were obtained in `shift-and-drift' mode in which the azimuth and elevation of the telescope were fixed, and the telescope therefore scanned in RA at a fixed declination as the Earth rotated. At the frequency we consider here, RFI is mainly terrestrial in nature and there is more gain to be made by minimising exposure to RFI by drift scanning, than can be made in the reduction of $1/f$ noise by scanning the telescope at rates higher than sidereal.
The location of the fields observed at Parkes is shown in Fig.~\ref{footprints} and Table~\ref{observation}, and the key observational parameters are listed in Table~\ref{MPIPAF}.

\begin{table}
	\centering
	\caption{WiggleZ survey fields Observed using the Parkes telescope with the PAF.}
	\label{observation}
	\begin{tabular}{lccr}  
		\hline
		Field name & Total integration time & Calibrators\\
		\hline
		3h & 22.6h & PKS~B1934-638, Hydra A\\
		1h & 8.7h & PKS~B1934-638\\
		0h & 15.3h & PKS~B1934-638, Hydra A\\
	    22h & 13.3h & PKS~B1934-638\\
		\hline
	\end{tabular}
\end{table} 

\begin{table}
	\centering
	\caption{Key observational parameters.}
	\label{MPIPAF}
	\begin{tabular}{lccr} 
		\hline
		Parameter & Value\\
		\hline
		Bandwidth & 384 MHz \\
		Central frequency & 891.5 MHz \\
		Spectral resolution & 18.5 kHz \\
        Number of channels & 20736 \\
        Cycle time & 4.5 s \\
        Polarisations & 2 \\
        Beams & 16 \\
		\hline
	\end{tabular}
\end{table}

\subsection{HI data reduction} 

\begin{figure} 
\centering
\includegraphics[width=0.75\textwidth]{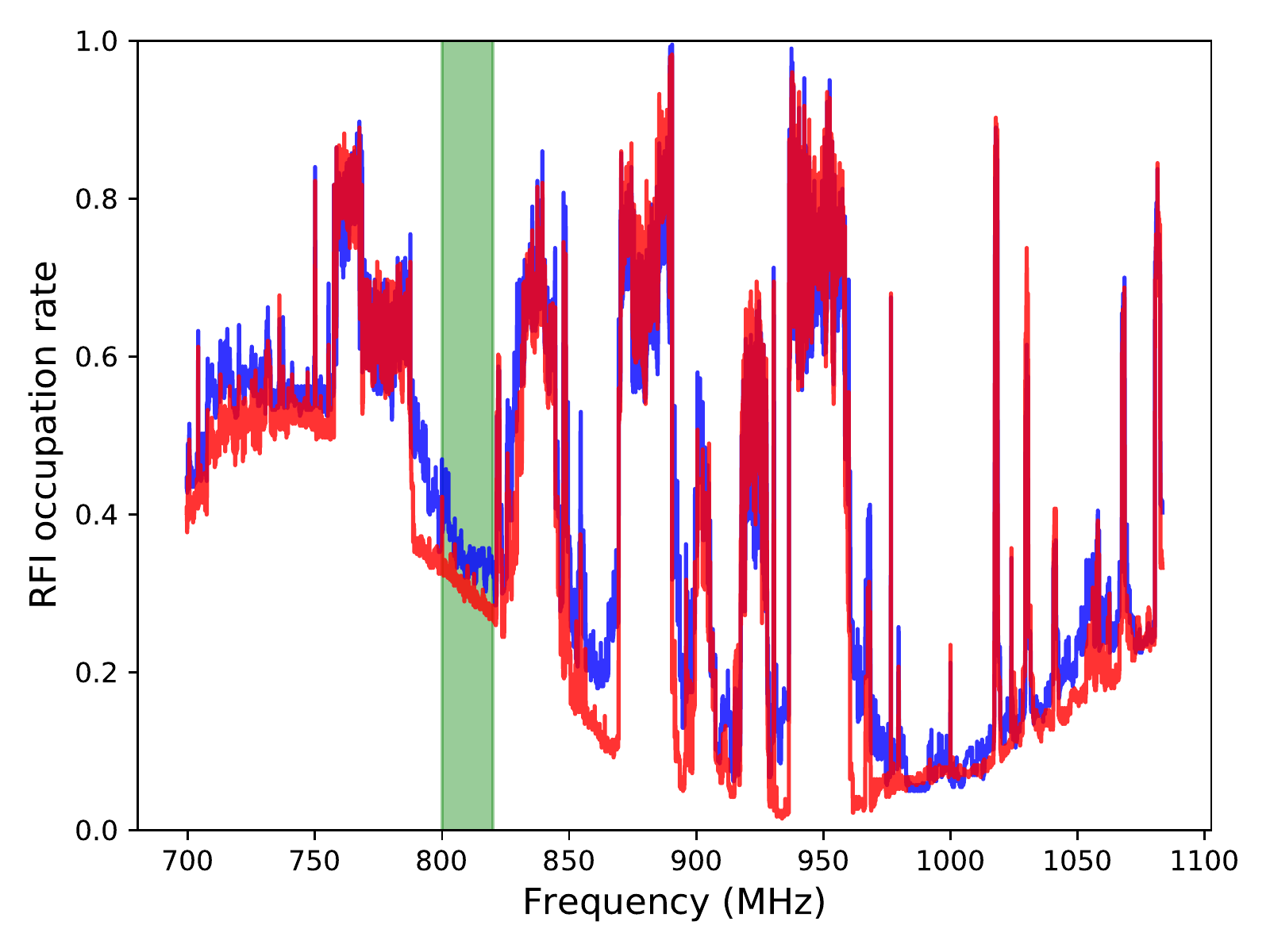}
    %\caption{An example of a calibrated system temperature as a function of frequency for the 16 beams in a scan. Red and blue colors denote the two polarisations.}
    \caption{The RFI/bad data occupancy for the two orthogonal polarisations (shown in red and blue). An occupancy of 1 means that no useful data was available. The green shaded area denotes the frequency range 800 to 820 MHz ($z=0.73-0.78$) analysed in this paper.}
    \label{tsys}
\end{figure}

%Fig.~\ref{tsys} shows an example of the calibrated system temperature measurement during the observations. Some frequency bands were masked because of severe RFI contamination. 
%Fig.~\ref{tsys} shows an example of the calibrated system temperature measurement during the observations. Some frequency bands were masked because of severe RFI contamination.
Fig.~\ref{tsys} shows the RFI occupancy in the whole frequency range, calculated from the fraction of the data at a given frequency where the system temperature was higher than 600 K during calibrator observations.
In this paper, we use the RFI-free frequency range of 800 to 820 MHz corresponding to a central redshift of $z=0.75$. Subsequent papers will consider other redshift ranges.

The data reduction procedure is as follows: 
\begin{enumerate}
\item We converted the raw data from arbitrary units to units of Kelvin. To do this, we calculated the scaling factors using on-source and off-source spectra for the calibrators, scaling the raw data in the targeted fields with the corresponding frequency-dependent factors. These scaling factors were measured to have uncertainties of $\sim11\%$. 

\item For each scan and each frequency channel, we calculated the running median for $\pm 30$ integration cycles, and subtracted the lowest median value from that channel. 
In the selected frequency band, the system temperatures of some beams were also very high, up to $\sim800$~K, due to bad beam-forming calibration.
We discarded beams with a median system temperature higher than 500 K.  

\item We then subtracted smooth foreground emission and RFI in an iterative way. Given that the foreground is mainly made up of Galactic synchrotron radiation and free-free emission with a monotonic variation with frequency, we removed a linear fit to the spectral data for each integration cycle.
Before going to the next step, we iteratively masked 3$\sigma$ outliers 
until the result converged.

\item To remove long-term temporal variations, we removed a linear fit to the data in each spectral channel for all scans.
Then we iteratively masked 3$\sigma$ outliers until the result converged. 

\item We calculated how much data had been flagged in each cycle, each channel and each scan. Cycles or channels with high masked fractions compared to the average were completely flagged. 
Where cycles or channels were bracketed with masked data on either side, the mask was dilated to also mask these cycles or channels. 
We again iteratively masked 3$\sigma$ outliers until the result converged. 

\item The calibration procedures (iii) $-$ (v) were repeated twice. 
\end{enumerate}

\begin{figure} 
	\includegraphics[width=\textwidth]{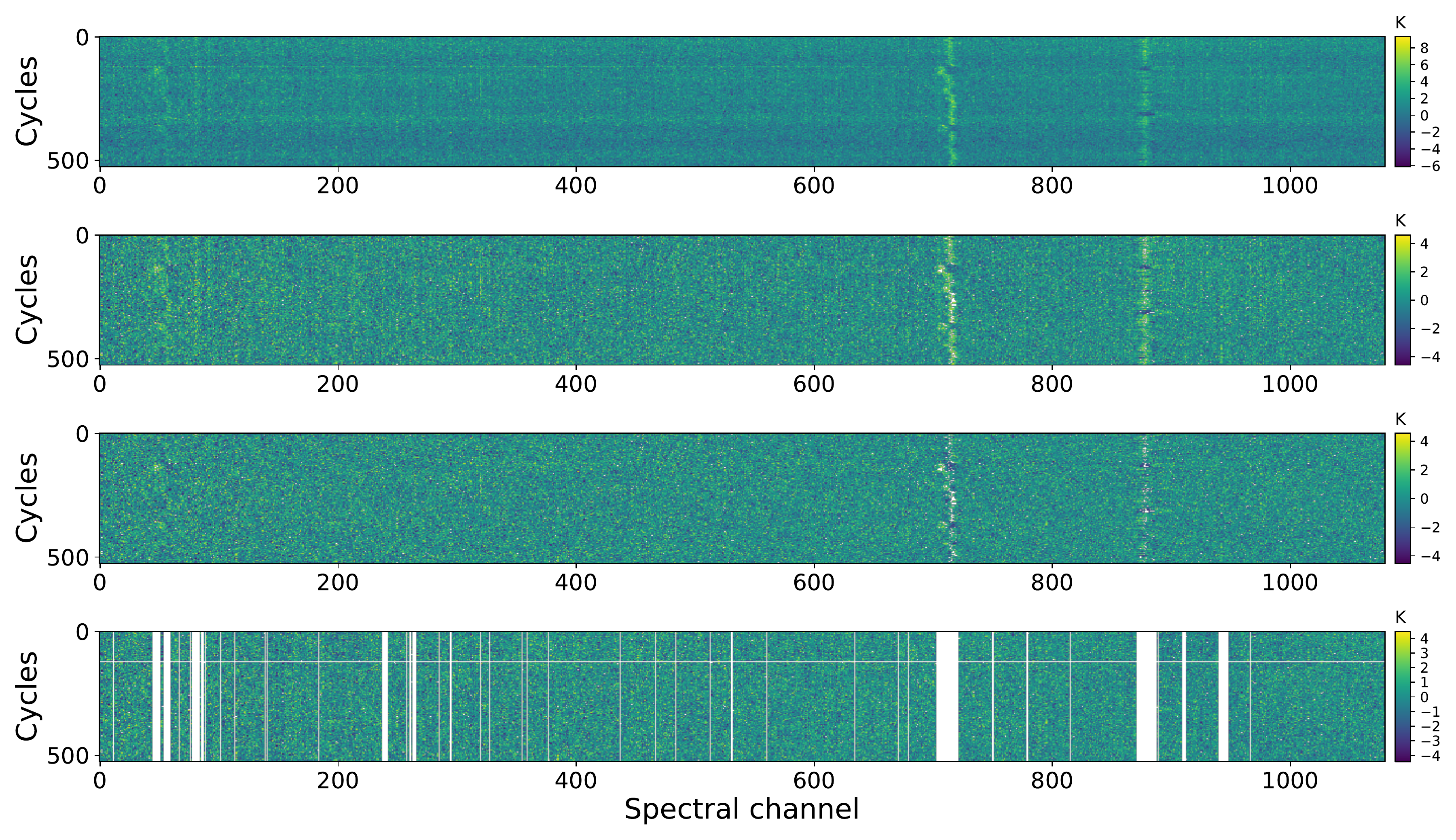}
    \caption{An example of the calibration and flagging process. Top panel: a waterfall frequency-time plot of the raw data containing RFI, receiver noise and foreground signal. Second panel: data with slow-moving spectral structure removed. Third panel:  data with slow-moving temporal structure and RFI removed.
Bottom panel: final masked data.}
    \label{calibration}
\end{figure}

Fig.~\ref{calibration} shows an example of the different steps in the calibration process, with the bottom panel showing an example of the final calibrated data. 

\begin{table}
	\centering
	\caption{Key parameters of the gridded data cubes and simulations in the WiggleZ 3 hr, 1 hr, 0 hr and 22 hr fields.} 
	\label{dimensions} 
\begin{tabular}{cclcc} 
\hline
\multicolumn{1}{l}{} & \multicolumn{1}{c}{} & \multicolumn{1}{c}{Gridded area (J2000)} & \multicolumn{1}{c}{Volume (Mpc$^3$)} & \multicolumn{1}{c}{Dimensions} \\
\hline
\multirow{4}{*}{observation} & 3 hr field  &$02^{\rm h} 53^{\rm m} < $ RA $ < 03^{\rm h} 28^{\rm m}$, $-18^\circ 36^\prime < $ Dec $ < -05^\circ 42^\prime$ & $603\times 412 \times 124 $ & (105, 104, 1080)\\
 & 1 hr field &$00^{\rm h} 30^{\rm m} < $ RA $ < 01^{\rm h} 05^{\rm m}$, $00^\circ 59^\prime < $ Dec $ < 05^\circ 12^\prime$ &  $200\times 408 \times 124 $   & (46, 94, 1080)\\
& 0 hr field &$23^{\rm h} 21^{\rm m} < $ RA $ < 23^{\rm h} 58^{\rm m}$, $-09^\circ 00^\prime  < $ Dec $ < 01^\circ 24^\prime$ & $486\times 425 \times 124$  & (112, 98, 1080)\\
 & 22 hr field & $21^{\rm h} 22^{\rm m} < $ RA $ < 22^{\rm h} 00^{\rm m}$, $-05^\circ 00^\prime < $ Dec $ < 04^\circ 42^\prime$ & $456\times 451 \times 124 $ & (139, 95, 1080)\\    
\hline
\multirow{4}{*}{simulation}  & 3 hr field & $02^{\rm h} 08^{\rm m} < $ RA $ < 04^{\rm h} 13^{\rm m}$ , $-19^\circ 21^\prime < $ Dec $ < -04^\circ 57^\prime$& $668\times 477 \times 124 $ & (120, 119, 1080)                               \\
& 1 hr field &$23^{\rm h} 45^{\rm m} < $ RA $ < 01^{\rm h} 50^{\rm m}$, $00^\circ 14^\prime < $ Dec $ < 05^\circ 57^\prime$& $265\times 473 \times 124 $& (61, 109, 1080)                              \\
 & 0 hr field &$22^{\rm h} 36^{\rm m} < $ RA $ < 00^{\rm h} 43^{\rm m}$, $-09^\circ 45^\prime  < $ Dec $ < 02^\circ 09^\prime$ & $556\times 495 \times 124 $ &  (128, 114, 1080)                              \\
 & 22 hr field &$20^{\rm h} 37^{\rm m} < $ RA $ < 22^{\rm h} 45^{\rm m}$, $-05^\circ 45^\prime < $ Dec $ < 05^\circ 27^\prime$ & $521\times 516 \times 124 $ & (154, 110, 1080)     \\
\hline
\end{tabular}
\end{table} 

The key parameters of the final gridded data cubes are summarised in the first row of Table~\ref{dimensions}.
Grid points were computed using the weighted median of all recorded brightness temperature values (all beams, all times, both linear polarisations) within a radius of half a beam around each given grid point. 
The weights were computed as the product of two parameters: (1) the reciprocal of the product of rms noise levels in the channel and cycle containing the observed value; and (2) the normalised value of the Gaussian at the position of the observation, where the Gaussian is centred on the grid point and is of the same width as the beam. 
The angular spacing is a quarter of the beam size, $5.58^\prime$ per pixel, 
with a frequency spacing of 18.5 kHz, corresponding to a comoving transverse size of 4.3 $h^{-1}$ Mpc and a comoving line-of-sight distance of 0.11 $h^{-1}$ Mpc at the central frequency channel.

\begin{figure} 
	\includegraphics[width=\textwidth]{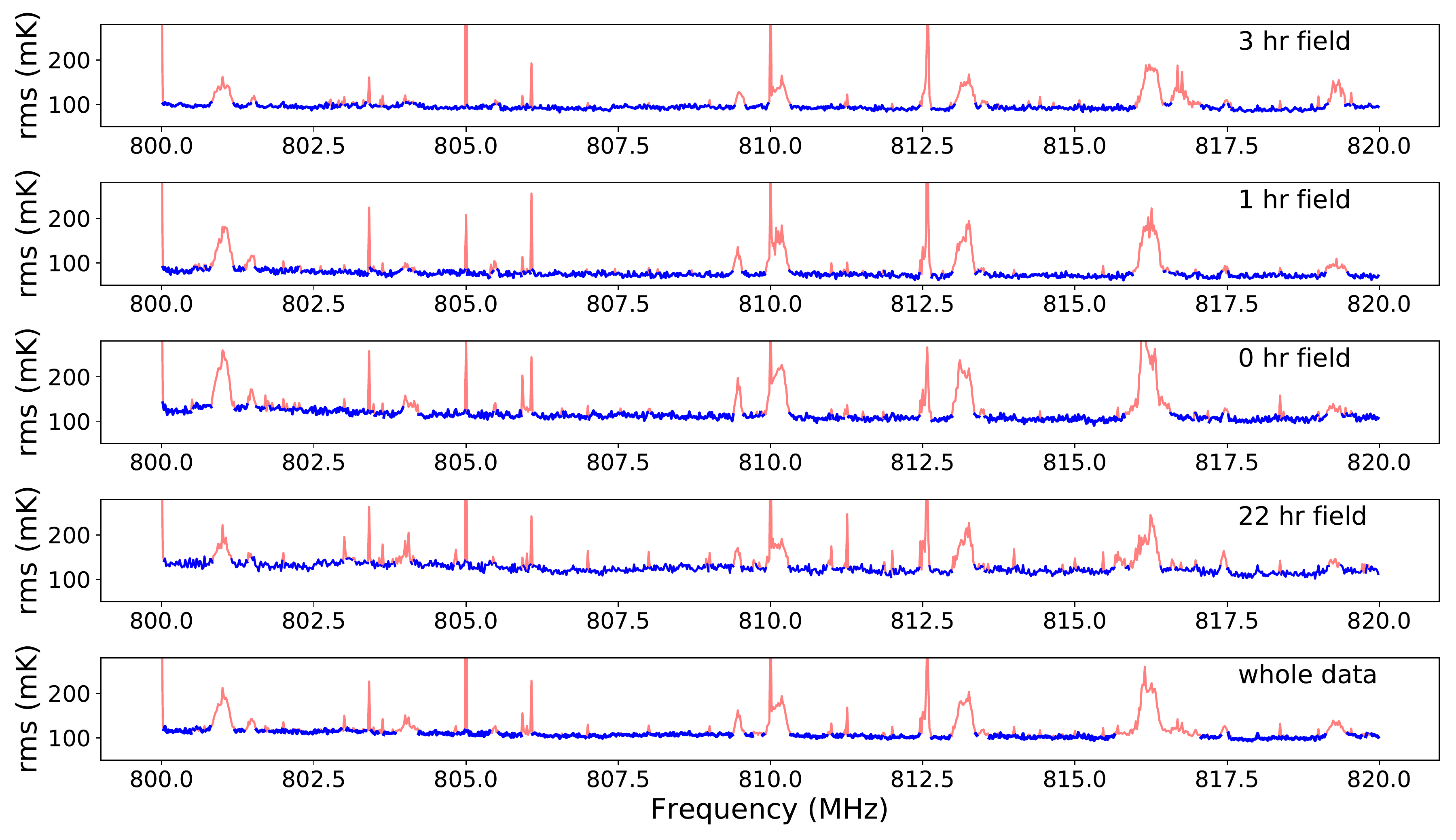} 
    \caption{The rms noise level in the gridded data cubes for each field over the frequency range considered in this paper. The frequency channels in red were masked. The frequency channels in blue correspond to the final data cube. The variation in the base rms levels between the four fields reflects the different system temperatures in these fields.}
    \label{flagrfi}
\end{figure}

Finally, the rms noise level was calculated for each channel in the gridded data cubes.  
In each field we iteratively fit a third-order polynomial to the rms spectrum and masked channels with an rms deviating by more than 3$\sigma$.
As RFI appears in almost the same frequency channels in all fields, we merged the masks from all four fields and applied to the individual fields.
Fig.\ref{flagrfi} shows the corresponding spectral masks for the individual fields, and the resultant merged mask.

\begin{figure} 
	\includegraphics[width=\textwidth]{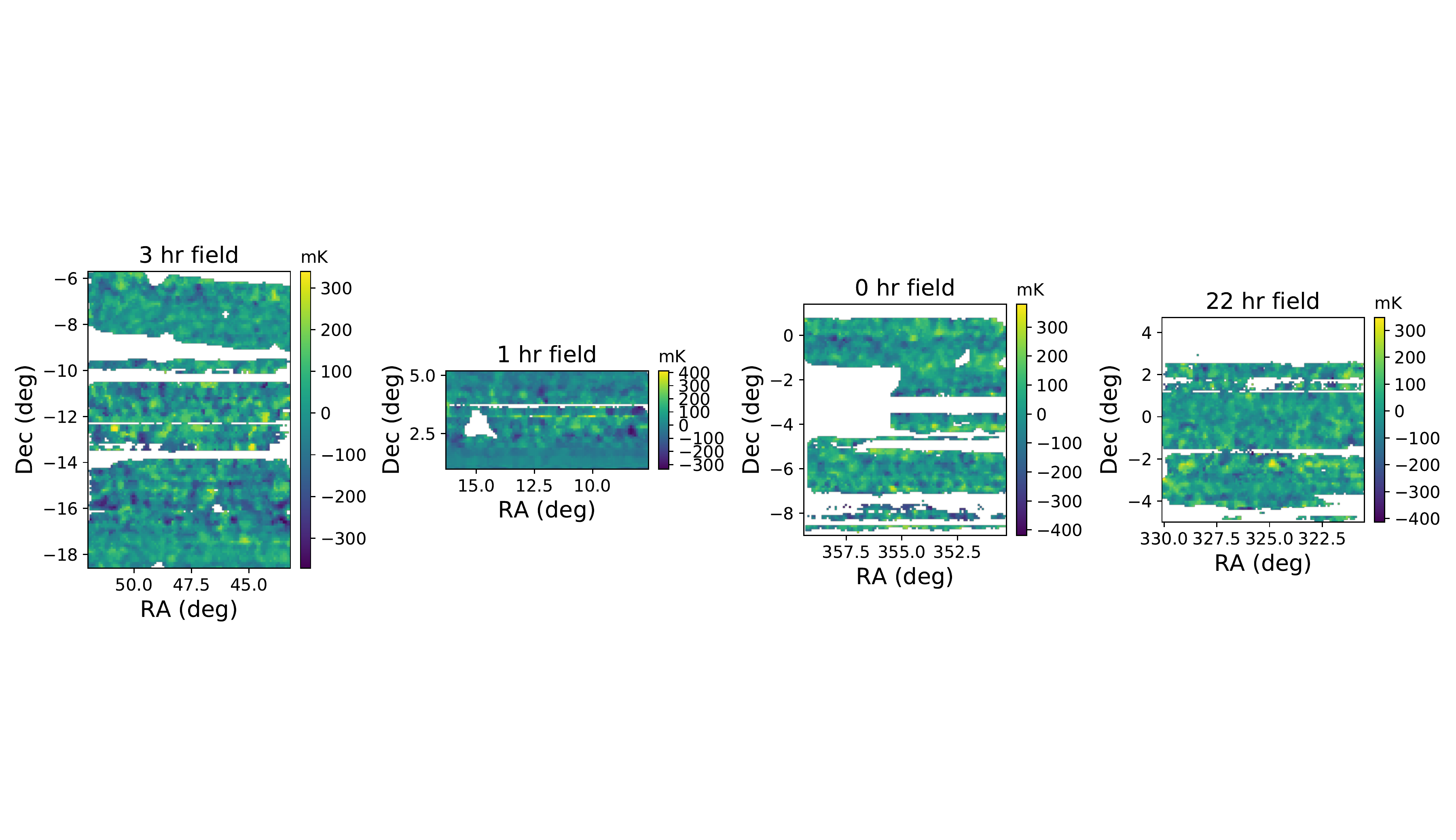}
    \caption{The final gridded HI maps for the four WiggleZ fields at the central spectral channel. White areas are masked due to either high rms noise levels or lack of optical data.}
    \label{HI_maps}
\end{figure}

The resultant data cube also has different noise levels in different directions. 
We calculated the rms noise level for each masked spectrum in the data cubes and further masked 22.0\% of the highest rms in the areas with WiggleZ coverage (see Appendix). 
The resultant cubes were down-sampled by averaging every 20 channels, resulting in a rms noise level of 35.5 mK for the whole data set.
Fig.~\ref{HI_maps} shows the final gridded {\HI} maps we obtained in the four fields at the central frequency channel.

\subsection{Optical data}
The optical catalog we used for cross-correlation is from the WiggleZ Dark Energy survey which measured redshifts for more than 240,000 galaxies up to redshift $z \sim 1$ in seven fields with a total area of approximately 1000 deg$^2$. Full details of the design and calibration are given in \citet{WigglZ}.

The optical density fields are derived from the corresponding WiggleZ catalogs from the final data release \citep{Drinkwater18}, which have 1446, 361, 733 and 1407 galaxies within the redshift range of the 3 hr, 1 hr, 0 hr and 22 hr regions, respectively. 
Due to that the optical sample are {\it GALEX}-selected so in some areas near bright stars it had no input data. 

We gridded the optical data into cubes, representing each galaxy by a 3-dimensional Gaussian with a full width at half maximum of 150 km~s$^{-1}$ in velocity space and 22\farcm3 in angular extent (to match the radio data). The units for this spectrum are arbitrary because only the optical overdensity field is required. The exact shape of the spectrum is also arbitrary in that the WiggleZ galaxies are only used as tracers of the underlying density field. 

To allow for WiggleZ selection effects, we generated 1000 random optical catalogues for which the selection functions and completeness are maintained while positions and redshifts of galaxies are randomised.
We computed these randomised optical density fields in the same way and then the real optical density field is corrected as follows:
\begin{equation}
	\centering
    \delta_{\textrm{opt}} = D/R\times \langle R\rangle/\langle D\rangle - 1 
	\label{del_opt}   
\end{equation}
where $D$ is the original computed optical density field and $R$ is the average density field from random catalogues at the same position. 
To avoid error introduced by dividing with small $R$ value, we then masked directions with $R<$ 1, which excludes $\sim 8$ per cent of all directions.

\begin{figure} 
	\includegraphics[width=\textwidth]{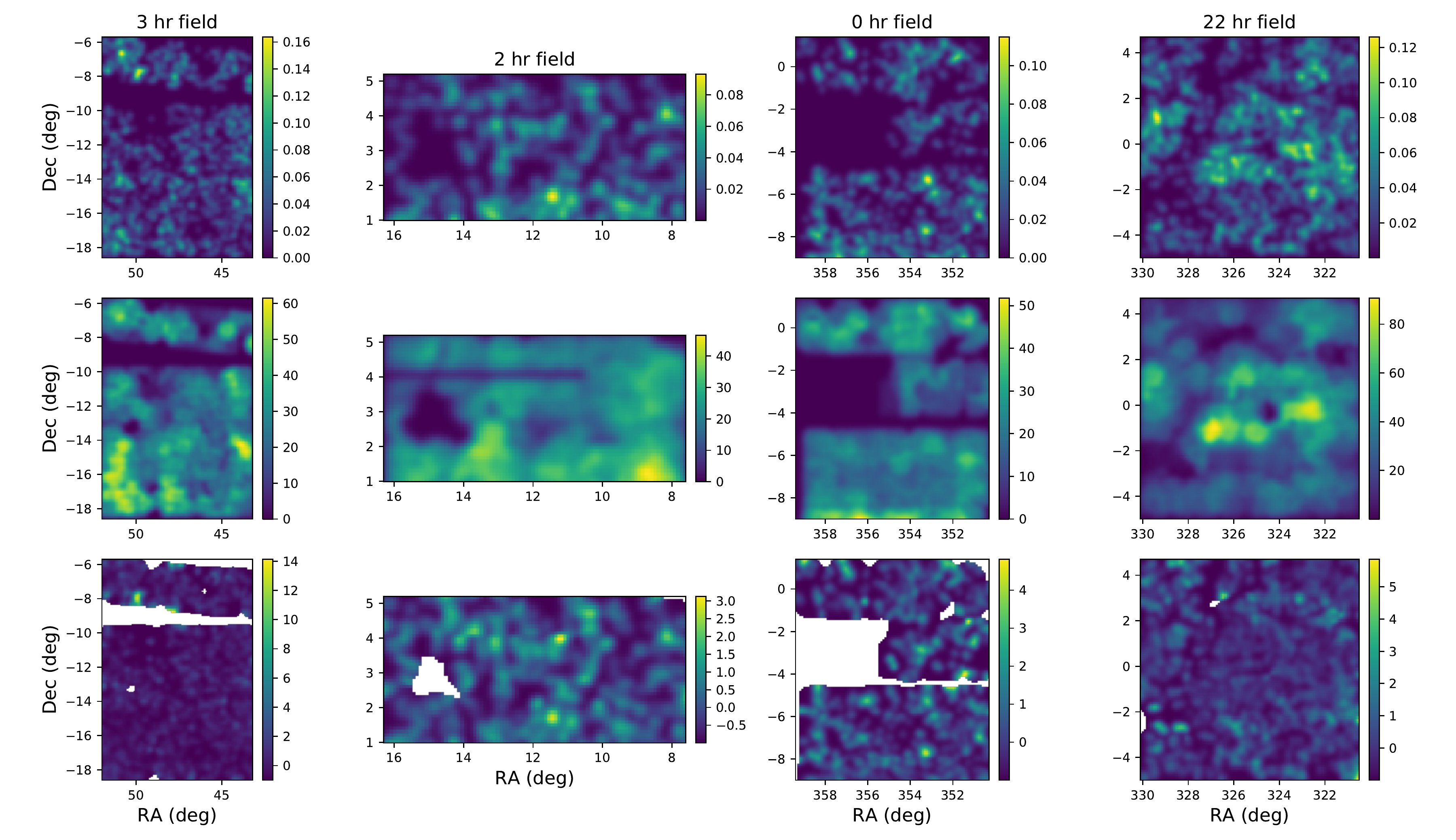} 
    \caption{Upper panels: the projected optical density fields computed from the WiggleZ catalogue.
    Middle panels: corresponding images of the ratio $\langle R\rangle/\langle D\rangle$ used to correct for the selection function. 
    Bottom panels: the final corrected and masked projected optical density fields.}
    \label{opt_maps}
\end{figure}

Fig.~\ref{opt_maps} shows the original optical density fields, the $\langle R \rangle/\langle D\rangle $ used for correction and the final corrected and masked optical density fields in four fields.

\section{Simulations}
\label{s:sims}
To understand the effect of our pipeline on the observed signal, we simulated an IM survey.
In our simulation, we assume that the distributions of neutral hydrogen and dark matter can be represented by the power spectrum generated using the CAMB package\footnote{https://camb.info/readme.html} \citep{Lewis00}.  
To create a simulated data cube, we first created the 3-D matter distribution in Fourier space. 
The values of the voxels in this data cube are complex numbers, the square of their modulus being related to the value of the power spectrum.

To cover the whole gridded area, we set the size of each simulated field to be 1.5 degrees larger than the real observation along both the RA and Dec directions, and with a spectral resolution the same as the gridded HI cubes. 
After inverse Fourier transform, the comoving volumes of the generated neutral hydrogen density fields are larger than the sizes of the real \HI\ data cubes previously discussed. 
The lower part of Table~\ref{dimensions} summarises the key parameters of the simulations in the four fields.

We convolved the simulated density field with the Gaussian PAF beam pattern and assigned coordinates to each voxel.
We then conducted a mock observation of the simulated density field using the actual scanning strategy, assuming identical system temperatures to the real data, and applied our gridding and masking algorithm to the mock data files. 

\begin{figure} 
	\includegraphics[width=\textwidth]{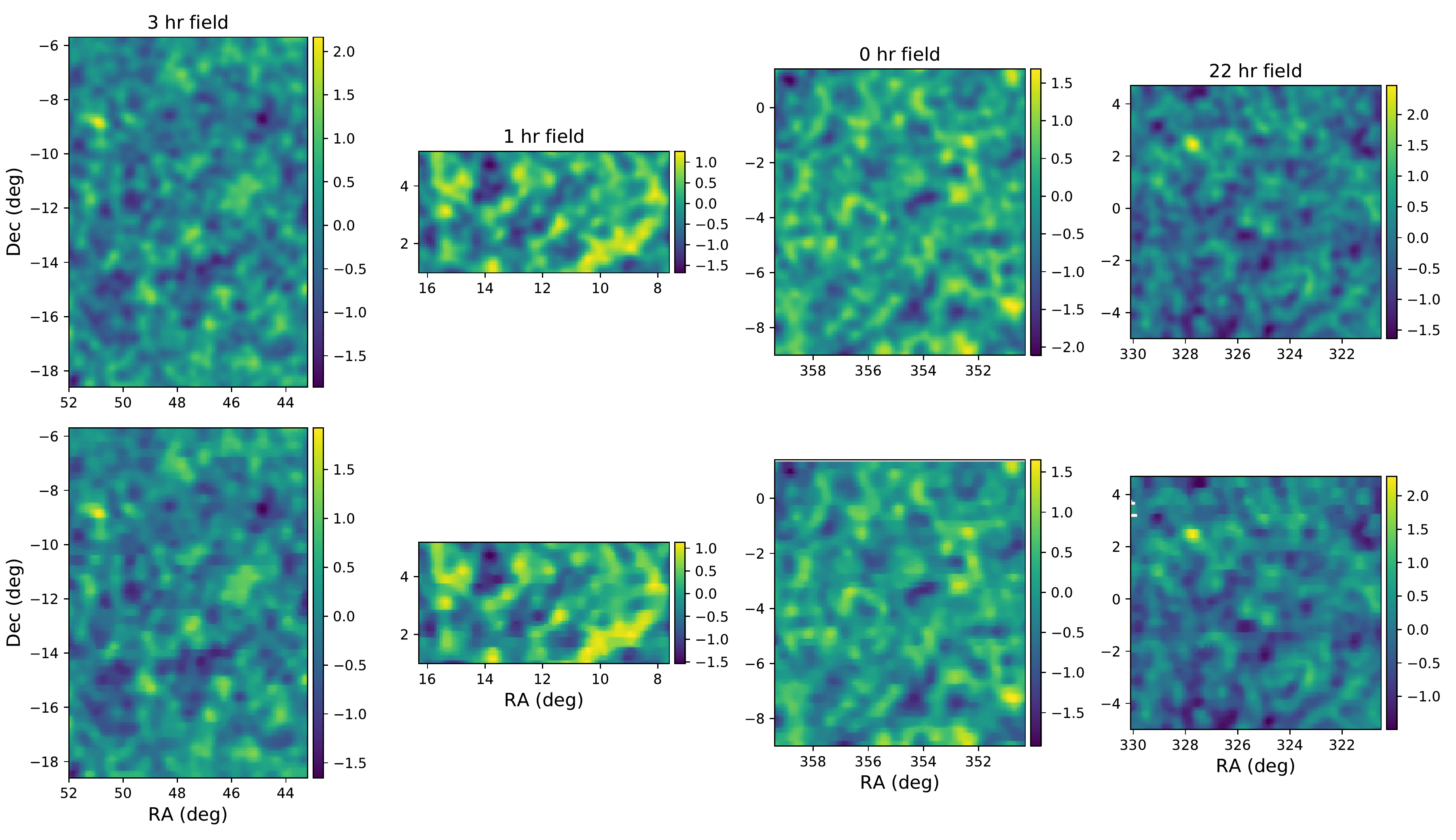}
    \caption{An example plane of a simulated data cube for the 3 hr, 1 hr, 0 hr and 22 hr fields.
    Upper panels: the original signal-only density field convolved with the beam.
    Lower panels: the observed density field in a mock survey.}
    \label{mockobs}
\end{figure}

Fig.~\ref{mockobs} shows an example of a single plane of the convolved simulated density field and the corresponding mock observation at the central frequency channel. 
We cross-correlated both, and compared it with the auto-correlation of the former. The result shows that the observed correlations are 6.3\%, 8.1\%, 6.7\% and 6.8\% smaller than the true correlations in the 3 hr, 1 hr, 0 hr and 22 hr field respectively, demonstrating the effect of the observing procedure and pipeline distortions on the signal strength.
For the whole masked dataset, an approximate average correction of 7.4\% is adopted.

\section{Residual foreground contamination}

\begin{figure}  
\centering
\includegraphics[width=0.6\textwidth]{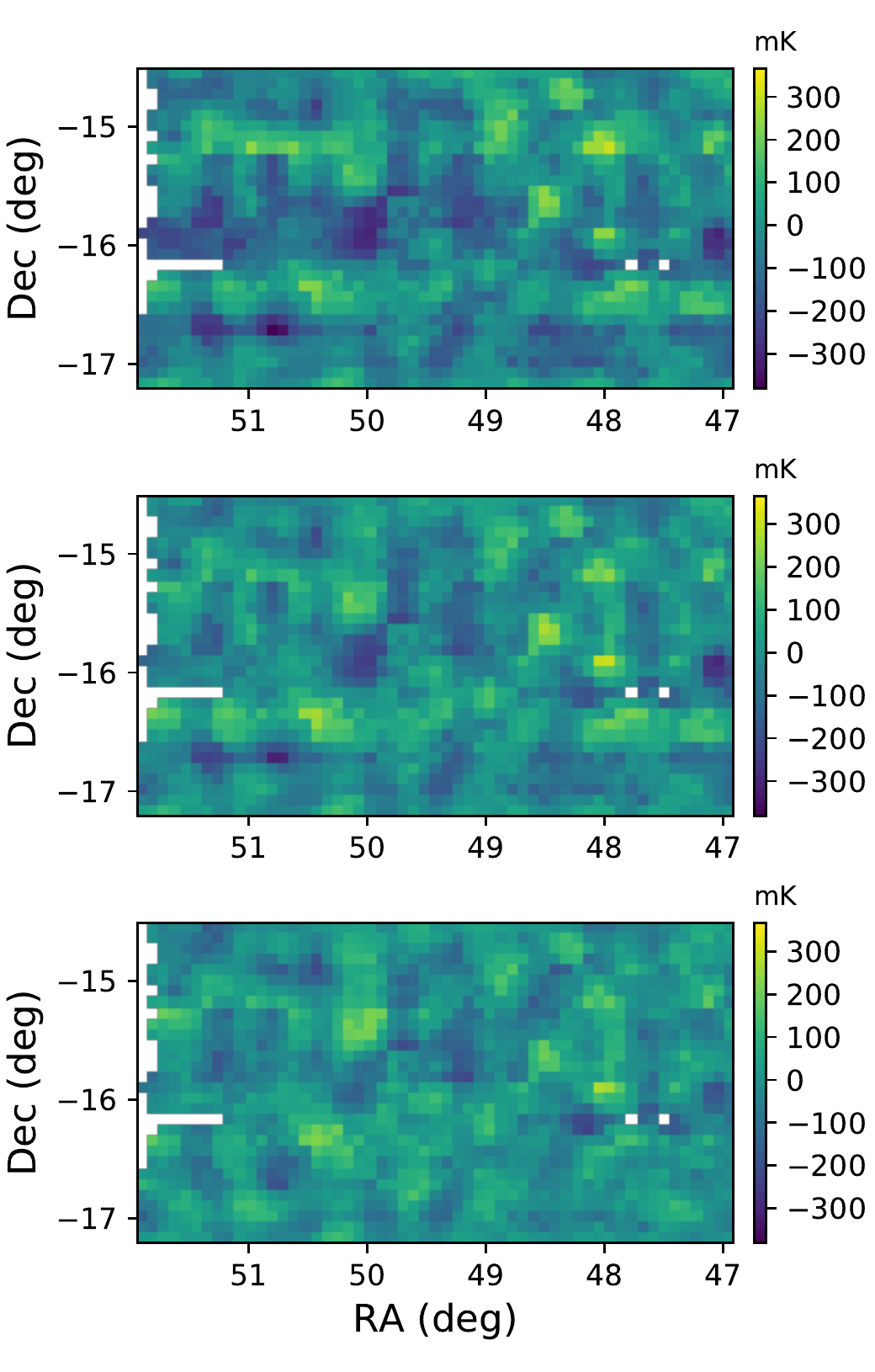}
    \caption{From top to bottom: an expanded view of an {\HI} data plane at the central channel in the 3-hr field  after the removal of 0, 30 and 120 modes, respectively.}
    \label{svdex}
\end{figure}

Despite careful attention to flagging and calibration, residual foreground and RFI contamination remain in the data at a level that is higher than the \HI\ signal we seek to measure. We therefore applied the technique of singular value decomposition (SVD) to our data cube, one of several foreground removal techniques applied by other authors to IM data sets \citep{Chang10,Masui2013,Anderson18}. SVD is a generalization of standard eigenvalue decomposition techniques which can be applied to non-square matrices. It decomposes the data cube into orthogonal modes of decreasing amplitude, with the low-order modes mainly containing the unwanted foreground and RFI residuals. 
In the SVD process, the 3D {\HI} data cubes are first transformed to four 2D matrixes, and then are concatenated to a $m\times n$ matrix $M$, where $m$ is the total pixel number on the plane perpendicular to line-of-sight and $n$ is the channel number. The decomposition of matrix $M$ is given by:
%\begin{center}
\begin{equation}
	\centering
    M = U\Sigma V^* ,
	\label{svd}   
\end{equation}
%\end{center} 
where $U$ is the $m\times m$ matrix of left-singular vectors, $\Sigma$ is an $m\times n$ matrix where the singular values are the non-negative reals on the diagonal, and $V^*$ is the conjugate transpose of $V$, which is the $n\times n$ matrix of right-singular vectors. The singular values of $M$ denote the amplitudes of each mode after decomposition. 
As the amplitudes of foreground residuals are usually assumed to be higher than the \HI\ signal, the low-order SVD modes are expected to mainly consist of these residuals, which makes possible a `cleaner' cross-correlation with the optical density field once these modes are removed and the \HI\ data cube is reconstructed from $M$.
Fig.~\ref{svdex} show examples of the recovered zoom-in imaged {\HI} data plane in the 3-hr field at the central channel after the removal of 0, 30 and 120 modes, respectively.

\begin{figure} 	
\includegraphics[width=\textwidth]{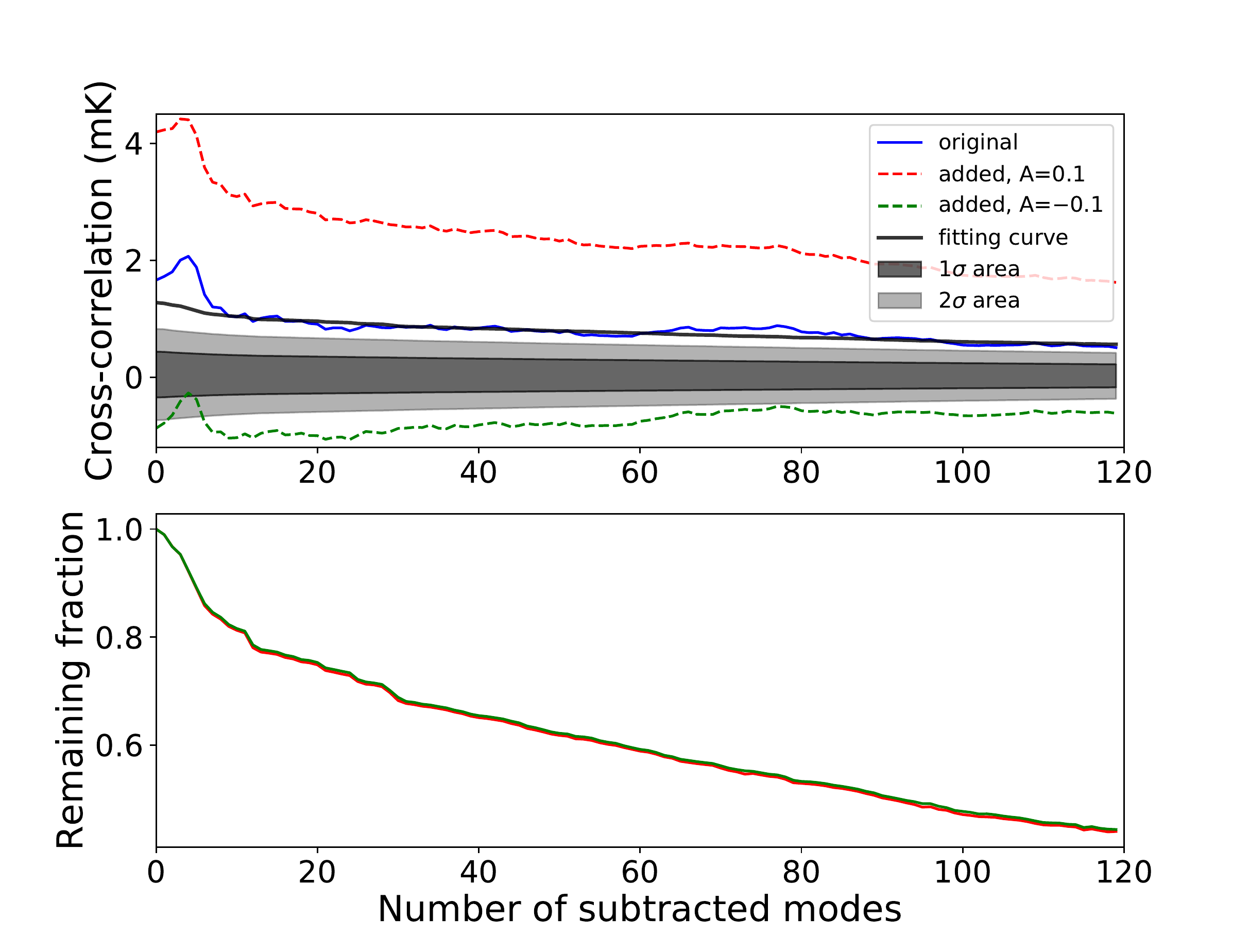}
     \caption{Cross-correlation of the whole data set. Upper panel: the cross-correlation of the density fields as a function of the number of subtracted SVD modes $N$. The blue solid line is the cross-correlation of the \HI\ and optical density fields. The red and green dashed line are the correlations resulting from the addition of the optical density field to the \HI\ density field, with $A$=0.1 and $-0.1$, respectively. The black solid line results from fitting the signal-loss function in the lower panel to the observed cross-correlation signal in blue. Dark and light shades shows the 1-$\sigma$ and 2-$\sigma$ confidence areas generated using 100 random optical catalogues.
     Lower panel: the fraction of the cross-correlation signal which remains intact following the SVD foreground subtraction as a function of $N$. The green and red (overlapping) curves correspond to results from the corresponding green and red dashed lines in the upper panel, respectively. 
     }
    \label{svdloss}
\end{figure}

\section{Results}

The cross-correlation of the \HI\ and optical data is shown in Fig.~\ref{svdloss}  as a function of the number $N$ of SVD modes removed. The fluctuations in the cross-correlation are large at small $N$ where the foreground and RFI components are dominant. At large $N$ the cross-correlation signal becomes low due to the SVD process removing genuine \HI\ signal.

In order to estimate this signal loss, we combined the optical density field with the \HI\ cube using a scale factor $A$, replacing $\Delta T_\textrm{b}$ with $\Delta T_\textrm{b} + A\delta_{\textrm{opt}}$ in the cross-correlation.
As shown in Fig.~\ref{svdloss}, this results in an increased/decreased cross-correlation amplitude which, as expected, also diminishes with $N$. The loss factor as a function of $N$ can be derived from the ratio of these measurements, as shown in the lower panel of Fig.~\ref{svdloss}. 

The ratio of the measured cross-correlation values to the loss factor therefore gives $N$ estimates of corrected cross-correlation signal. Rather than selecting an arbitrary value for $N$, we have simply used a median fit of the loss curve to the measured signal, as shown by black solid line in the upper panel of Fig.~\ref{svdloss}. This results in a median cross-correlation signal. 

To account for signal loss due to the data reduction and gridding process (see Section~\ref{s:sims}), we apply a 7 per cent correction and derive our final result of $\left\langle\Delta T_\textrm{b}\delta_\textrm{opt}\right\rangle = 1.32\pm 0.42$ mK. The significance of this result is at the 3-$\sigma$ level, so is a tentative detection. 

The errors represent
%The 97.7 per cent confidence upper limit is 2.2 mK.
the standard deviation of the cross-correlation derived from 100 random optical catalogues in which the same optical selection function as the WiggleZ survey is applied, while the positions and redshifts have been randomised. The error is therefore purely statistical.

To estimate errors arising from our processing and analysis pipeline, we tested the pipeline with different parameters, including the width of the velocity convolution function, the upper limit for $N$ used in fitting the correlation amplitudes, and the fitting method itself. With reasonable changes, these parameters only cause minor changes (within 10\%) to the final results.

However, other more major sources of errors not accounted for include cosmic variance and the discrete sampling of the optical density field. The sparseness of the WiggleZ data in particular may be a major limitation in accurate calculation of the optical density field. The simulations required to model this are outside the scope of this paper.

In principal, the detected signal can be related to the cosmic \HI\ density {\OHI} using the following relation:

\begin{equation}
\begin{split}
    \langle\Delta T_\textrm{b}\delta_{\textrm{opt}}\rangle= 44  b r \langle\delta^2_{\textrm{opt}}\rangle\Bigg(\frac{{\OHI}h}{2.45\times10^{-4}}\Bigg) 
    \frac{(1+z)^2}{E(z)} ~\mu\textrm{K} 
	\label{omega_HI}
\end{split}
\end{equation}
\citep{Chang10}, where $\Delta T_\textrm{b}$ is the \HI\ brightness temperature fluctuation, $\delta_{\textrm{opt}}$ is the normalized optical density field, $b=\langle \delta_{\HIsub}^2\rangle^{0.5}\big/ \langle \delta_{\textrm{opt}}^2\rangle^{0.5}$ is the bias factor, $r=\langle \delta_{\HIsub}\delta_{\textrm{opt}}\rangle\big/ \big(\langle \delta_{\HIsub}^2\rangle\langle\delta_{\textrm{opt}}^2\rangle \big)^{0.5}$ is the stochasticity, and $E(z)=H(z)/H_{0}$ and $h$ is the Hubble constant in units of 100 km s$^{-1}$ Mpc$^{-1}$. However, as the terms in the above are model-dependent, and as there are potential large uncertainties in correcting for the (sparse) optical density field noted above, realistic galaxy simulations are required, so this derivation is deferred to a later paper.

\section{Summary and Future Work}
We have conducted an \HI\ intensity mapping experiment with the Parkes 64-m telescope in four equatorial fields which have redshift data out to $z\sim 1$ from the WiggleZ redshift survey. The \HI\ data were taken with a novel phased array feed (PAF) which permitted simultaneous observations over 16 separate beams. This paper presents the data reduction procedure in a redshift band at $z\sim 0.75$ which was judged clear of radio frequency interference. 

The rms noise level in the data cubes in the range 800--820 MHz was 35.5 mK at a spectral resolution of 0.37 MHz. 
After foreground removal using a SVD technique, the cross-correlation signal with the WiggleZ optical density fields in the same redshift range was detected at close to the 3$\sigma$ level. 
After correction for signal loss caused by the data reduction and analysis pipeline, the amplitude of the observed cross-correlation is $\left\langle\Delta T_\textrm{b}\delta_\textrm{opt}\right\rangle = 1.32\pm 0.42$ mK (statistical errors).
%, corresponding to a 97.7 percent confidence upper limit of 2.2 mK.

Despite the presence of high system temperature and strong RFI, we have been able to make useful IM measurements at a cosmological redshift with a PAF. 
The interpretation of the cosmic \HI\ density that corresponds to the measured cross-correlation signal (or upper limit) requires  a more sophisticated simulation that is beyond the scope of this paper. However, preliminary indications are that a future generation cryogenic PAF will be able to substantially improve on the accuracy of these observations and, with more attention to foreground contamination, may even permit a detection of the auto-correlation signal, thus dispensing with the requirement for a matching optical redshift survey.

\normalem
\begin{acknowledgements}
We thank Chris Blake and Yi-Chao Li for their constructive discussion and suggestions. LL thanks the Chinese Scholarship Council and University of Western Australia for the financial support. We thank the staff of the Parkes radio telescope for their technical assistance. The Parkes radio telescope is part of the Australia Telescope National Facility which is funded by the Commonwealth of Australia for operation as a National Facility managed by CSIRO. Parts of this research were conducted by the Australian Research Council Centre of Excellence for All-sky Astrophysics (CAASTRO),
through project number CE110001020 and the Australian Research Council Centre of Excellence for All Sky Astrophysics in 3 Dimensions (ASTRO 3D), through project number CE170100013. This work was also supported by the National Key $R\&D$ Programme of China under grant number 2018YFA0404603. 
\end{acknowledgements}
  
\bibliographystyle{raa}
\bibliography{references}

\appendix
\section{Masking pointing directions}
The cross-correlation which we aim to measure can be expressed as: 
\begin{equation}
	\centering
    C=\langle \delta_{\textrm{opt}}\Delta T_{\textrm{HI},\textrm{o}}\rangle = \langle \delta_{\textrm{opt}} \Delta T_{\textrm{HI},\textrm{r}}\rangle+\langle \delta_{\textrm{opt}}T_\textrm{n}\rangle
	\label{none}   
\end{equation} 
where $\Delta T_{\textrm{HI},\textrm{o}}$, $\Delta T_{\textrm{HI},\textrm{r}}$ and $T_\textrm{n}$ are the observed \HI\ brightness temperature, real \HI\ brightness temperature and the noise, respectively. 
As the expectation values $E(\delta_{\textrm{opt}})=0$,  $E(T_\textrm{n})=0$ and $T_\textrm{n}$ is uncorrelated with $\delta_{\textrm{opt}}$, the variance of $C$ can be computed as: 

\begin{equation}
\begin{split}
    D(C)= D(\delta_{\textrm{opt}} \Delta T_{\textrm{HI},\textrm{r}})+ D( \delta_{\textrm{opt}}T_\textrm{n}) \\
     = D(\delta_{\textrm{opt}} \Delta T_{\textrm{HI},\textrm{r}})+D( \delta_{\textrm{opt}})D(T_\textrm{n}).
	\label{dc}   
\end{split}
\end{equation} 
The cross-correlation is calculated by averaging from all voxels:  $\sum C_j/N$, where $N$ is the number of voxels used for averaging, and the subscript $j$ denotes the voxel number. The expected variance of the mean estimate of $C$, assuming unweighted averaging and a universal $D(\delta_{\textrm{opt}})$, can therefore be approximated as:

\begin{equation}
\begin{split}
\frac{m}{N^2}\sum_{j=1}^{N}D_j=\frac{m}{N^2}\sum_{j=1}^{N}D_j(\delta_{\textrm{opt}} \Delta T_{\textrm{HI},\textrm{r}}) +  \frac{m}{N^2}\sum_{j=1}^{N}D_j(\delta_{\textrm{opt}})D_j(T_\textrm{n}) \\
=\frac{m}{N^2}\sum_{j=1}^{N}D_j(\delta_{\textrm{opt}} \Delta T_{\textrm{HI},\textrm{r}}) +  D(\delta_{\textrm{opt}})\frac{m}{N^2}\sum_{j=1}^{N}D_j(T_\textrm{n}) .
\label{fracn} 
\end{split}
\end{equation}
where $m$ is the number of pixels per area. The first two terms on the right side of this equation are from the real density fields and cannot be reduced. So the uncertainty may be expected to reach a minimum when the final term $S = \sum D_j(T_\textrm{n})/N^2$ is minimised.
In the observation $\Delta T_\textrm{n}>>\Delta T_{\textrm{HI},\textrm{r}}$ and final gridded data quality depends much more on the pointing direction rather than frequency. We therefore compute the noise variance for each spectrum, $D_j(T_\textrm{n})$ and calculate $S=\sum D_i(T_\textrm{n})/N^2$.
 
\begin{figure}
 \centering
 \includegraphics[width=0.75\textwidth]{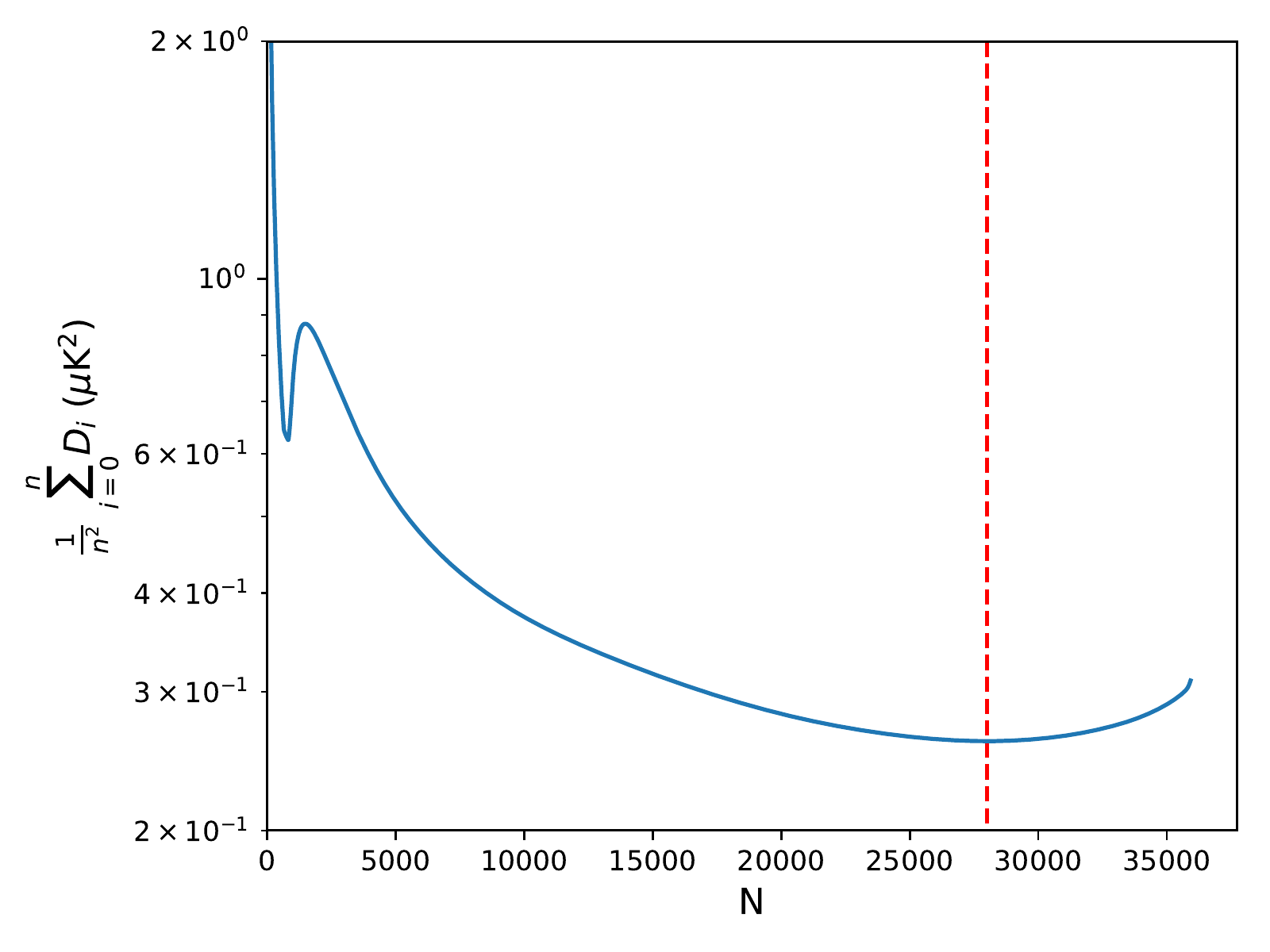}
     \caption{$\sum_{i=1}^{N}D_i/N^2$ as a function of the number of used pixel directions, $N$, in our data. The red dashed line denotes the minimal value of the blue curve as the cut for masking pixel directions in our data. }
    \label{varcut}
\end{figure}

In keeping with our preference to mask poor quality data, rather than applying continuous weights, we now eliminate directions with the highest variance. The appropriate metric is to calculate the value for $N$ for which the above quantity $S=\sum D_i/N^2$ is minimised after sorting the $D_i$ in ascending order. 
Fig.~\ref{varcut} shows $S$ a function of $N$ for our data. The result shows that $S$ reaches a minimum when 22.0\% of the directions with the highest rms noise level are masked. 
 
\end{document}